# NEUROPEPTIDE Y AND ITS INVOLVEMENT IN CHRONIC PAIN

Marta Diaz-delCastillo[a], David P.D. Woldbye[b], Anne Marie Heegaard[a]

Corresponding author: Marta Diaz-delCastillo (marta.castillo@sund.ku.dk)
Phone number: (+45) 71 83 26 07
Postal address: Jagtvej 160, Copenhagen Ø, 2100 (Denmark).

[a] Department of Drug Design and Pharmacology, Faculty of Health and Medical Sciences, University of Copenhagen. Jagtvej 160, Copenhagen Ø, 2100 (Denmark). amhe@sund.ku.dk

[b] Department of Neuroscience, Faculty of Health and Medical Sciences, University of Copenhagen. Blegdamsvej 3, Copenhagen N, 2300 (Denmark). woldbye@sund.ku.dk

## Abbreviations

CNS, Central nervous system; NPY, Neuropeptide Y; PNS, Peripheral nervous system; PYY, Peptide YY; PP, Pancreatic polypeptide; CPON, C-flanking peptide of NPY; DRG, Dorsal root ganglia; Y1R, Y1 receptor; Y2R, Y2 receptor; Y5R, Y5 receptor; SP, Substance P; NPY-LI, Neuropeptide Y-like immunoreactivity; CFA, Complete Freund´s Adjuvant; GABA, γ-aminobutyric acid.

## Abstract

Chronic pain is a serious condition that significantly impairs the quality of life, affecting an estimate of 1.5 billion people worldwide. Despite the physiological, emotional and financial burden of chronic pain, there is still a lack of efficient treatments. Neuropeptide Y (NPY) is a highly conserved endogenous peptide in the central and peripheral nervous system of all mammals, which has been implicated in both pro- and antinociceptive effects. NPY is expressed in the superficial laminae of the dorsal horn of the spinal cord, where it appears to mediate its antinociceptive actions via the Y1 and Y2 receptors. Intrathecal administration of NPY in animal models of neuropathic, inflammatory or post-operative pain has been shown to cause analgesia, even though its exact mechanisms are still unclear. It remains to be seen whether these promising central antinociceptive effects of NPY can be transferred into a future treatment for chronic pain.

# Introduction

Chronic pain has a higher prevalence than diabetes, coronary heart disease, stroke and cancer all together (Institute of Medicine Committee on Advancing Pain Research and Education, 2011) and the World Health Organization has estimated that 22% of patients seeking primary health care worldwide complain about chronic pain (Gureje et al., 1998). Several studies conducted in United States, Sweden, Denmark or United Kingdom show that chronic pain conditions cost billions of dollars due to a reduced level of productivity and a high rate of absenteeism, on top of the direct health care costs of long lasting, often chronic, treatments (Phillips, 2009). It has been estimated that 1 million working days per year were lost in Denmark, which has a population of 5.8 million inhabitants (Eriksen et al., 2006). Denmark has the highest level of non-recreational opioid consumption due to chronic pain (Eriksen et al., 2006), and more than 67% of the prescribed opioids are used for treatment of non-cancer related chronic pain (Eriksen et al., 2006, Phillips, 2009). What is even more important is that chronic pain greatly impairs the quality of life of many patients, who cannot lead a normal life (Iadarola and Caudle, 1997). However, the molecular mechanisms underlying the maintenance and development of chronic pain are still not fully elucidated.

Since the isolation of morphine from opium in 1800 (Klockgether-Radke, 2002), the main treatment for chronic pain has been the use of both natural and synthetically manufactured opioids targeting the opioid receptors in the central nervous system (CNS). However, this treatment lacks long-term efficacy, and it is accompanied by many side effects that contribute to the reduced quality of life of chronic pain patients (Kalso et al., 2004). The severity of these side effects and risk of abuse has led to opioids not being considered the first drugs of choice for the treatment of chronic pain (Finnerup et al., 2015). Furthermore, recent data suggest that, in the long run, opioids themselves may exacerbate chronic pain by acting at kappa- and mu-opioid receptors on microglial cells, promoting the release of cytokines and other pro-inflammatory molecules that contribute to the development and maintenance of pain (reviewed in (Giron et al., 2015)).

Traditionally, chronic pain has been classified as *inflammatory* and *neuropathic* pain, depending on whether it is caused by chronic inflammatory states (e.g. rheumatoid arthritis) or from lesions within the nervous system, respectively. In both cases, the pain state is characterized by persistent changes within the spinal cord and the pain modulation centers of the brain, as a consequence of the long-lasting peripheral changes in the activation of nociceptors. After depolarization of Aδ and C fibers that transmit the nociceptive signal to the dorsal horn of the spinal cord, the neuronal input is carried on towards supraspinal structures; the ascending spinothalamic tracts transmit then the nociception to the thalamus, from where it is directed to other brain areas related to the management of the pain sensation (reviewed in (Tracey, 2017)).

The dorsal horn of the spinal cord is known to be a key area in pain gating and modulation, and many studies have addressed the role of neuropeptides expressed in this region (e.g. calcitonin gene-related peptide (CGRP), substance P, somatostatin and enkephalins) and their involvement in nociceptive signaling (Marvizón et al., 2009, McCoy et al., 2013, Shi et al., 2014). Moreover, during the last two decades, neuropeptide Y (NPY) has appeared as another important component in nociceptive signaling modulation through its action on several receptors in the central nervous system, especially Y1R and Y2R. In spite of extensive research, the exact role of NPY in processing pain sensation is still not fully elucidated. The aim of this review is to give a comprehensive summary of the different studies addressing the role of NPY in pain modulation at the level of the spinal cord and dorsal root ganglia.

## Neuropeptide Y and its receptors

### Neuropeptide Y

Neuropeptide Y (NPY) is a 36 amino-acid peptide first isolated and sequenced from porcine brain in 1982, which is widely expressed in the CNS and peripheral nervous system (PNS) of mammals (Tatemoto, 1982). This neuroactive peptide, together with peptide YY (PYY) and pancreatic polypeptide (PP), with which it shares 70% and 50% amino acid sequence homology, respectively, constitute the so-called NPY family of peptides (Dumont et al., 1992). The sequence similarity of the three molecules is translated into a common U-shaped, hairpin-like, tertiary structure consisting of an amphiphilic α-helix connected through a β-turn to a polyproline-like type II helix. This tridimensional structure defines the affinity and efficacy with which the peptides bind to their many receptors (Cerda-Reverter and Larhammar, 2000).

The biosynthesis of NPY involves the translation of a pre-pro-NPY molecule that is directly translocated into the endoplasmic reticulum, where its signal peptide is removed. The precursor pro-NPY is then cleaved into NPY and C-terminal peptide of NPY (CPON) by prohormone convertases, and the NPY peptide suffers two subsequent truncations (by a carboxypeptidase and a peptidylglycine alpha-amidating monooxygenase) in order to reach the mature and biologically active form of NPY. This final molecule is characterized by its C-terminal amidation, which prevents its degradation by carboxypeptidases (Silva et al., 2002).

Consistent with its wide distribution in the CNS and PNS, NPY has been implicated in many biological effects and brain disorders, including energy homeostasis, obesity and anorexia (Loh et al., 2015), anxiety (Heilig et al., 1993, Sørensen et al., 2004), epilepsy (Baraban et al., 1997, Woldbye et al., 1997), depression (Farzi et al., 2015), drug addiction (Gonçalves et al., 2015), opioid withdrawal (Woldbye et al., 1998 , Clausen et al., 2001, Sørensen et al., 2004), itch (Bourane et al., 2015) and pain (Brumovsky et al., 2007, Arcourt et al., 2017). Interestingly, most of these disorders are characterized by the sharing of behavioral and psychological components. In this review, we will focus on NPY expression and function in the dorsal root ganglia and spinal cord in relation to pain.

## NPY receptors

Up to date, five different receptors of NPY have been cloned in mammals: Y1, Y2, Y4, Y5 and y6 (Herzog et al., 1992, Larhammar et al., 1992, Gerald et al., 1995, Rose et al., 1995, Gerald et al., 1996, Gregor et al., 1996). Despite y6 being an active, functional receptor in mice and rabbits, it is absent in rats, and its functionality is lost in humans due to a single base-par deletion (Gregor et al., 1996). Both NPY and PYY present a similar Y-receptor binding profile (Lin et al., 2004b); however, the distribution pattern of each receptor is different.

All NPY receptors are $G_{i/o}$ protein coupled, and thereby they exert their effects by promoting conformational changes in associated proteins that lead to the activation of several molecular pathways. For example, the mitogen-activated protein kinase (MAPK) pathway can be triggered upon NPY binding, leading to the inhibition of adenylyl cyclase activity (Brumovsky et al., 2007) and, therefore, a decrease in the levels of cAMP. Actually, it is not necessarily the binding of NPY to its receptor that directly triggers this pathway, but rather the indirect transactivation of the insulin growth factor receptor (IGFR) which leads to the activation of ERK1/2 (Lecat et al., 2015). This, in addition to the modulation of the inositol 1,4,5-triphosphate ($IP_3$) pathway, results in modifications of the intracellular calcium concentration of the neuron and, therefore, changes in its membrane potential (Brumovsky et al., 2007). Furthermore, NPY binding to Y1R can directly affect intracellular $Ca^{2+}$ concentration by activating or inhibiting neuronal N- and P/Q-type $Ca^{2+}$ channels (Wang, 2005) and R- and L-type $Ca^{2+}$ channels, as well as $K^+$ channels (Xiong and Cheung, 1995, Sun et al., 2001), while its action on Y2R is limited to the modulation of the N- and P/Q-type $Ca^{2+}$ channels (McCullough et al., 1998, Wang, 2005).

## Role of NPY and its receptors in pain processing in the dorsal root ganglia and spinal cord

In the dorsal root ganglia (DRG), NPY-like immunoreactivity (NPY-LI) is present at very low levels, with the exception of very few small primary afferent neurons (Wakisaka et al., 1992, Magnussen et al., 2015). Nevertheless, NPY expression after peripheral nerve injury is prominently up-regulated in the large and medium diameter cell bodies of the primary sensory neurons innervating the dorsal horn of the spinal cord; a phenomenon that has been described by several research groups over the years (Wakisaka et al., 1992, Noguchi et al., 1993, Sten Shi et al., 1999, Benoliel et al., 2001, Hökfelt et al., 2007).The novel synthesis of NPY can be explained as an adaptive response to the hyperalgesia-induced excitatory signaling (Munglani et al., 1995).

Both Y1R and Y2R are expressed in CGRP-positive (presumably nociceptive) neurons located in the rat DRG, suggesting a role in nociception mediated by NPY (Zhang et al., 1994, Zhang et al., 1997), since they provide a physical circuit that allows the autocrine or paracrine effect of NPY on pain modulation (Verge et al., 2002). Y1R is located in small and some medium size neurons (Zhang et al., 1994, Landry et al., 2000, Brumovsky et al., 2004), while Y2R are mostly found in the medium and large neuronal somas (Zhang et al., 1997, Landry et al., 2000). Using an optogenetic approach to target Y2R expressing neurons, Arcourt et al. further demonstrated that the Y2R is likely to be expressed in a subset of peptidergic A-fiber nociceptors activated by mechanical stimuli in the noxious range (Arcourt et al., 2017).

Furthermore, it has been described that up to 40% of the NPY-positive neurons found in the DRG following mechanical nerve injury, also express Y1R and Y2R (Landry et al., 2000). However, while Y2R mRNA in the cell bodies of sensory neurons is increased after peripheral axotomy (Zhang et al., 1997, Landry et al., 2000), Y1R mRNA levels in the DRGs decrease (Brumovsky et al., 2004).

At the spinal level, around 15% of the inhibitory interneurons of lamina I-III in the rat are NPY-immunoreactive and have been shown to co-localize with GABA (Polgár et al., 2010). A more recent study using *NPY::Cre* transgenic mice identified an abundant population of NPY-expressing inhibitory interneurons in laminae III/IV, and, to a lesser extent, in laminae I and II. In addition, the study showed a decrease in the number of NPY-expressing interneurons postnatally (Bourane et al., 2015). Many attempts have been made to classify these interneurons with regards to their morphology. However, a study using transgenic mice (Iwagaki et al., 2015) indicates that NPY is equally present in several differently sized inhibitory interneurons in the dorsal horn of the spinal cord. This study explored the morphology of NPY-expressing cells in the dorsal horn of mice for the first time, allowing its comparison with other well-defined neuronal subpopulations. The authors combined electrophysiology and immunohistochemistry to investigate cells expressing GFP under the control of an NPY promoter. This way, they concluded that while other neuronal populations can be classified on the basis of their somatodendritic anatomy, the NPY-GFP$^{+}$ cells are morphologically heterogeneous (although never islet cells) and hence cannot be classified according to their morphology.

As shown in figure 1, following peripheral nerve damage, an increase in NPY-LI can be observed in the middle regions of laminae I-IV in the dorsal horn of the spinal cord (Brumovsky et al., 2004). Similarly, peripheral inflammation (in the rat Complete Freund´s Adjuvant (CFA) model) induces an increase both in NPY mRNA transcripts and NPY-LI in the ipsilateral dorsal horn of the spinal cord, particularly in lamina II (Ji et al., 1994). Furthermore, a combination of sciatic nerve transection with dorsal rhizotomy suggests that the increase in central NPY could have both a spinal and ganglionic origin: while dorsal rhizotomy does not affect the NPY-LI in laminae I and II, it leads to decreased NPY-LI in laminae III and IV (Ohara et al., 1994). Additional studies using a similar approach (sciatic nerve injury followed by dorsal rhozotomy) suggest that, upon synthesis of NPY in the DRGs, the peptide is transported to the axon terminals of sensory neurons in the spinal cord (Ossipov et al., 2002)

Differential expression of the NPY receptors is found in the dorsal horn of the spinal cord. Y1Rs are highly expressed in laminae I-V (in addition to area X and the ventral horns), where up to seven different populations of Y1+ neurons have been described, suggesting a complex role in nociception (Brumovsky et al., 2006). These neuronal populations include a few primary afferents and many excitatory interneurons (Y1+, SOM+, V-GLUT2+) that can be hyperpolarized by NPY (Brumovsky et al., 2006). On the other hand, Y2R expression is restricted to laminae I and II (Brumovsky et al., 2005), or even predominantly to lamina II (Arcourt et al., 2017), reflecting the localization of Y2Rs in the central terminals of the primary afferents (Zhang et al., 1997, Brumovsky et al., 2005). In contrast to Y1Rs that are found both pre- and post-synaptically (Brumovsky et al., 2002, Brumovsky et al., 2006), the Y2Rs have never been described in dorsal horn interneurons. Thus, Y2Rs are considered to act primarily pre-synaptically, in accordance with the hypothesis of these receptors being centrifugally transported to the dorsal horn (rhizotomy completely abolishes spinal Y2R expression), especially after peripheral nerve injury or axotomy (Brumovsky et al., 2005).

## NPY-mediated analgesia

For several years it remained controversial whether NPY exerts pro- or anti-nociceptive actions. It seems clear now that the divergent effects of this peptide depend on its route of delivery (Cougnon et al., 1997, Lin et al., 2004a, Brumovsky et al., 2007). Thus, peripheral administration has been reported to be pronociceptive. For instance, subcutaneous injection of the peptide or a Y2R agonist following sciatic nerve injury exacerbates both mechanical and thermal hyperalgesia, while administration of a Y1R agonist in the same manner increases mechanical, but reverses thermal, hyperalgesia (Tracey et al., 1995). In this study, the authors suggest that the NPY receptors mediating the pronociceptive effect are located on sympathetic nerve terminals and that, following nerve injury, their expression, affinity or intracellular effect is amplified.

On the other hand, intrathecal administration of NPY triggers analgesia (Hua et al., 1991, Brumovsky et al., 2007, Hökfelt et al., 2007, Thomsen et al., 2007, Intondi et al., 2008). Already in 1991, intrathecally administered NPY was shown to induce analgesia after thermal stimulation (Hua et al., 1991). More recently, central administration of NPY has been shown to mediate antinociceptive effects in a number of chronic pain models, such as the sciatic nerve injury model of neuropathic pain (Intondi et al., 2008), the CFA model of inflammatory pain (Solway et al., 2011) or the plantar incision model of postoperative pain (Yalamuri et al., 2013). Particularly, the research of Solway et al. (Solway et al., 2011) highlights the importance of endogenous NPY in the transmission of the nociceptive signal. After conditional NPY knock-down, they demonstrated that NPY tonically inhibits neuropathic pain, as the knock-down of the peptide before or after peripheral injury was enough to restore cold and mechanical allodynia, suggesting that NPY exerts a continuous suppression of hyperalgesia in chronic pain states.

While several behavioral tests have also been developed to categorize, compare and quantify pain-related behaviors (Lariviere et al., 2002), another useful tool as a global molecular marker of nociception is the expression of the protein c-Fos. It is known that the proto-oncogene *c-fos* is an immediate early activation gene, rapidly expressed following noxious stimuli (Hunt et al., 1987). For that reason this peptide has widely been used for mapping neurons involved in the transmission of nociceptive stimuli (Menétrey et al., 1989, Coggeshall, 2005). Along with a reduction in noxious stimulus-induced behavior, intrathecal administration of NPY has been correlated with a 51% reduction of c-Fos expression in lamina I-VI of the dorsal horn (Intondi et al., 2008). The same study found a 56% reduction of c-Fos expression after an equivalent dose of morphine, which has long been known to reduce stimulus-induced c-Fos up-regulation. Intondi et al. also found that intrathecal injections of NPY reduced, in a dose-dependent manner, both formalin and spared nerve injury elicited allodynia and hyperalgesia, along with c-Fos expression in the dorsal horn, categorizing NPY as an "outstanding candidate for the interruption of spinal nociceptive transmission" (Intondi et al., 2008).

Many studies have addressed the involvement of the different NPY receptors in analgesia and a substantial amount of evidence points to Y1R as the main mediator of the anti-nociceptive effect of NPY. The crucial role of Y1R was confirmed by Naveilhan et al. (Naveilhan et al., 2001) when they created a mouse model with a targeted deletion of Y1R by homologous recombination. These mice presented hyperalgesia to acute thermal, cutaneous and visceral chemical stimulus, as well as mechanical hypersensitivity. From these results the authors concluded that this receptor is necessary for physiological and pharmacological central NPY-mediated analgesia.

While several spinal Y1R-positive neuronal populations have been identified, it is thought that the anti-nociceptive function of NPY is mainly exerted by a group of Y1R-expressing excitatory interneurons in the superficial laminae of the dorsal horn (Brumovsky et al., 2006, Hökfelt et al., 2007). These excitatory interneurons would be inhibited upon NPY binding to the Y1R. Wiley et al. studied the role of Y1R-expressing neurons in the dorsal horn of the spinal cord by permanently inhibiting them (Wiley et al., 2009). For that purpose, they intrathecally administered a conjugate of NPY and the ribosomal inactivating toxin saporin (NPY-sap), which destroyed Y1R-positive neurons in the dorsal horn without affecting the sensory neurons in the lumbar DRGs. The selective destruction of these interneurons reduced nociceptive reflexes such as formalin-induced and 44°C-hotplate related pain behaviors. This supports the hypothesis of Y1R being located on excitatory interneurons, which, when inhibited either by NPY or by their actual destruction, alleviate the nociceptive response. However, these results only provided evidence for the involvement of Y1R-expressing neurons in nociception, but not of their role in analgesia. Thus, Lemons et al. argued that in order to assess the role of NPY and Y1R in analgesia, it is necessary to verify that the nociceptive stimulus has been processed in the cerebrum (Lemons and Wiley, 2012). Using the same NPY-sap conjugate, they tested three different operant paradigms commonly used in the assessment of hypersensitivity and hyperalgesia in rats, before and after inducing CFA inflammation. Their results suggest that Y1Rs play a role in analgesia, since the loss of Y1R-positive neurons leads to an improvement on several escape tasks in relation to hot and cold stimuli and CFA-related allodynia.

One of the proposed mechanisms of Y1R-mediated anti-hyperalgesia is via inhibition of substance P (SP) release upon activation of the Y1R (Taylor et al., 2014). As it has been extensively studied, noxious or electrical stimulation of C-fibers promote SP release in the dorsal horn of the spinal cord (Yaksh et al., 1980, Adelson et al., 2009, Zhang et al., 2010). SP is then able to bind neurokinin 1 receptors (NK1Rs) in the dorsal horn neurons, which subsequently activate different nociceptive pathways, and several studies using NK1R antagonists, as well as total abolishment of the NK1R$^+$ neuronal population, have indicated a crucial role of SP in nociceptive transmission in the spinal cord (Radhakrishnan and Henry, 1991, Nichols et al., 1999). It has been shown that central *in vivo* NPY administration inhibits SP release (Duggan et al., 1991), and, more recently, Taylor et al. demonstrated that in animal models of inflammatory pain, there is an increase in the coupling of Y1R with G-proteins. In these models, intrathecal administration of NPY caused a reduction in NK1R internalization (Taylor et al., 2014). Taken together, these results support that the increased functionality of Y1R and its downstream signaling upon inflammation contributes to the inhibition of SP release from primary afferent terminals, causing a reduction in NK1R activation (Taylor et al., 2014).

However, it should be noted that there is a disagreement between animal and clinical trials in relation to the role of NK1R and SP in nociception. Several companies have evaluated the analgesic effect of NK1R agonists in phase II clinical trials, finding little (Dionne et al., 1998) or no effect of these agonists compared with placebo (reviewed in (Borsook et al., 2012)). There are several factors that could contribute to these differences, from the interspecies variation on NK1R localization in supraspinal sites, to the type of nociception, as animal testing heavily relies on measurements of evoked, rather than spontaneous pain (Hill, 2000).

Regarding the Y2R, several studies have indicated a role for these receptors in the processing of nociceptive signals. Spinal administration of a Y2R antagonist (BIIE0246) reversed the anti-nociceptive effect of NPY in the sciatic nerve injury model of neuropathic pain to the same extent as a Y1R antagonist (BIBO3304), suggesting that the effect of NPY is triggered upon binding to both receptors in the neuropathic pain model (Solway et al., 2011).

However, the involvement of Y2R in inflammatory pain remains unclear. In a traditional model of chronic inflammatory pain by unilateral plantar injection of CFA, intrathecal administration of NPY dose-dependently reduced the signs of allodynia and hyperalgesia, similarly to the neuropathic pain model. Again, intrathecal administration of the Y1R antagonist BIBO3304 reversed this behavior, but in this case the Y2R antagonist BIIE0246 showed no effect (Taiwo and Taylor, 2002). Nevertheless, recent studies demonstrate similar effects of both Y1R and Y2R agonists in models of neuropathic pain, as well as CFA (Solway et al., 2011). Thus, although NPY spinal analgesia via Y2Rs has been shown to modulate acute (Hua et al., 1991) and neuropathic pain (Xu et al., 1999), the involvement of Y2Rs in chronic inflammatory pain needs to be further elucidated.

The Y5 receptor has also been suggested to be involved in pain modulation (Thomsen et al., 2007), presumably via supraspinal mechanisms, and it remains to be studied whether this receptor plays a role at the spinal cord level.

What is clear to this date is that, in the spinal cord, Y1Rs are present in both primary afferents and spinal neurons (excitatory interneurons), whereas Y2R expression is restricted to primary afferents (Brumovsky et al., 2006). In general, Y1Rs are considered to have an inhibitory effect on excitatory interneurons (Brumovsky et al., 2006), while Y2Rs are thought to inhibit the release of excitatory neurotransmitters from the primary afferents (Brumovsky et al., 2005). This way, the inhibition of excitatory interneurons and projection neurons by NPYergic tonic expression in response to an insult could take place through both its action on Y1R or Y2R. Thus, NPY likely plays a homeostatic role in the spinal transmission of the nociceptive signal by binding to Y1Rs and Y2Rs (Solway et al., 2011, Taylor et al., 2014).

## NPY as a potential therapeutic target

Several factors have to be taken into account when considering the effect of centrally administered NPY in chronic pain. The heterogeneity of NPY receptors, the site of application and the different animal models of chronic pain can explain the diverse available information on NPY effects and mechanisms of action. However, several lines of evidence support that central NPY does produce analgesia and that its effect is mediated at the spinal (and supraspinal) level (Moran et al., 2004, Intondi et al., 2008, Taylor et al., 2014).

Taken together, NPY and its receptors Y1R and Y2R have been shown to play an important role in modulating chronic pain, and NPY receptors could be considered as drug targets for the development of new rationally designed drugs. However, there are several difficulties to overcome with this potential treatment target. The route of administration is one of them, since peripheral application of NPY agonists exacerbates pain-related behaviors, and it seems likely that only central administration of the peptide or drug formulations that allow the transport through the blood-brain barrier can have an analgesic effect.

Lumbar intrathecal delivery of drugs is associated with some discomfort and inconvenience for the patient that could only be overcome by high efficiency and long-lasting effects of the treatment, two main features that are still to be clarified in this particular case. Another obstacle for testing the clinical potential of NPY agonism in chronic pain conditions is the lack of small molecule non-peptidic NPY receptor agonists (Brothers and Wahlestedt, 2010, Mittapalli and Roberts, 2014). Thus, in this context, alternative strategies such as targeted delivery of the agonists using nanoparticles or gene therapy have been proposed (Chandrasekharan et al., 2013).

## Conclusion

Chronic pain is a highly prevalent condition for which there is a lack of effective treatments. Several studies have focused on the use of NPY for the treatment of this condition, showing that spinal administration of the peptide via intrathecal injections modulates the nociceptive stimulus and causes analgesia in animal models of chronic pain. The general belief is that NPY is synthetized *de novo* in the neurons of the dorsal root ganglia and their terminal regions in the spinal cord, where they inhibit the nociceptive pathway, serving as an adaptive compensatory mechanism in response to excessive excitatory signaling. The NPY receptors Y1R and Y2R are thought to mediate this effect, which would mainly take part in the dorsal horn of the spinal cord and constitute potential drug targets for the treatment of chronic pain. However, further research is needed to clarify the exact mechanism underlying the analgesic effect of NPY.

## Acknowledgements

This work was supported by the European Union´s Horizon 2020 Research and Innovation Program, under the Marie Sklodowska-Curie grant agreement No 642720.

Figures:

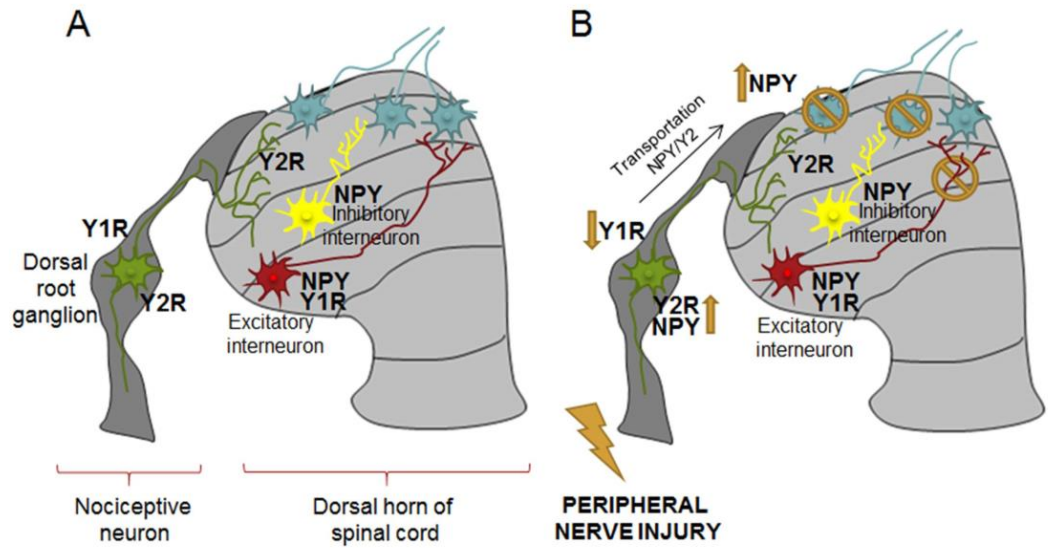

**FIGURE 1:** Red: excitatory interneuron. Yellow: inhibitory interneuron. Blue: Ascending spinothalamic tracts. **A**. Distribution of NPY, Y1R and Y2R in the spinal cord and dorsal root ganglia at the baseline situation (Magnussen et al., 2015, Arcourt et al., 2017). **B**. After peripheral nerve injury, the levels of Y1R in the dorsal root ganglia decrease (Brumovsky et al., 2004), while Y2R and NPY increase (Zhang et al., 1997, Landry et al., 2000). There is a transportation event of the peptide (Ossipov et al., 2002) and its Y2R (Ohara et al., 1994, Brumovsky et al., 2005) towards the dorsal horn, where Y2Rs are hypothesized to inhibit excitatory primary afferents (Brumovsky et al., 2005) while Y1R hyperpolarizes excitatory interneurons (Naveilhan et al., 2001, Brumovsky et al., 2006, Wiley et al., 2009).